\newcommand{\ba}{Ba$_2$CoGe$_2$O$_7$}
\newcommand{\bacu}{Ba$_2$CuGe$_2$O$_7$}
\newcommand{\bacuco}{Ba$_2($Co$_x$Cu$_{1-x})$Ge$_2$O$_7$}

\documentclass[aps,prb,twocolumn,groupedaddress]{revtex4}
\usepackage{graphicx}
\usepackage{bm}

\begin{document}
%\preprint{}

\title{Spin waves and the origin of commensurate magnetism in \ba.}
\author{A. Zheludev}
\email[]{zheludevai@ornl.gov}

\affiliation{Condensed Matter Sciences Division, Oak Ridge
National Laboratory, Oak Ridge, TN  37831-6393, USA.}

\author{T. Sato}
\author{T. Masuda\protect\footnote{Present
address: Condensed Matter Sciences Division, Oak Ridge National
Laboratory, Oak Ridge, TN  37831-6393, USA.}}
\author{K. Uchinokura} \affiliation{Department of Advanced Materials Science,
The University of Tokyo, Tokyo 113-8656, Japan.}

\author{G. Shirane}
\affiliation{Physics Department, Brookhaven National Laboratory,
Upton, NY 11973-5000, USA.}

\author{B. Roessli}
\affiliation{Laboratory for Neutron Scattering, ETH Zurich and
Paul Scherrer Institute, CH-5232 Villigen PSI, Switzerland.}

\begin{abstract}
The square-lattice antiferromagnet \ba\ is studied by means of
neutron diffraction and inelastic scattering. This material is
isostructural to the well-known Dzyaloshinskii-Moriya helimagnet
\bacu\, but exhibits commensurate long-range N\'{e}el order at low
temperatures. Measurements of the spin wave dispersion relation
reveal strong in-plane anisotropy that is the likely reason for
the suppression of helimagnetism.
\end{abstract}

\date{\today}
\maketitle

\section{Introduction}
Several years ago the square-lattice helimagnet \bacu\ was
recognized as an extremely interesting material for studying
Dzyaloshinskii-Moriya (DM) off-diagonal exchange interactions. A
great deal of attention was given to the incommensurate nature of
the magnetic ground state,\cite{Zheludev1996} a unique
field-induced incommensurate-to-commensurate (IC)
transition,\cite{ZheludevPRL1997,ZheludevPRB1998} and the field
dependence of the spiral spin structure.\cite{ZheludevPRB1997}
Studies of the spin wave spectrum in the
incommensurate\cite{ZheludevPRL1998,Zheludev1999} and
commensurate\cite{Zheludev1999} phases led to the first direct
observation of a new type of magnetic interactions in insulators,
the so-called KSEA
term.\cite{Moriya60,Kaplan1983,Shekhtman1992,Shekhtman1993,Entin1994}
Additional theoretical studies shed light on the nature of the
so-called ``intermediate phase''.\cite{Chovan2002}

A recent study indicated that an IC transition in \bacu\ can be
induced not only by applying an external magnetic field, but also
by a partial chemical substitution of the spin-carrying Cu$^{2+}$
ions by Co$^{2+}$.\cite{Sato2002} For all Co-concentrations $x$
the solid solution \bacuco\ orders magnetically at temperatures
between $T_N=3.2$~K ($x=0$) and $T_N=6.7$~K ($x=1$). Magnetization
data suggest that the helimagnetic state realized at $x=0$ gives
way to a canted weak-ferromagnetic structure at some critical
concentration $x_c$, estimated to be between $0.05$ and $0.1$. The
mechanism of this transition or crossover is poorly understood.
One possible explanation was proposed in
Ref.~\onlinecite{Sato2002}. The structure of \bacuco\ is
tetragonal, the $S=1/2$ Cu$^{2+}$ or $S=3/2$ Co$^{2+}$ ions
forming a square lattice within the $(a,b)$ crystallographic
plane. The dominant interaction is the antiferromagnetic (AF)
coupling $J$ between nearest-neighbor (NN) sites along the
$(1,1,0)$ direction. In the $x=0$ compound the helimagnetic
distortion is caused by the in-plane component $\mathbf{D}_{xy}$
of the Dzyaloshinskii vector $\mathbf{D}$ associated with the same
Cu-Cu bonds.\cite{ZheludevPRL1997,ZheludevPRB1998} This component
retains its direction from one bond to the next, and thus favors a
spin spiral state.\cite{Dzyaloshinskii1964} In contrast, the
out-of-plane component $\mathbf{D}_{z}$ is sign-alternating and
stabilizes a weak-ferromagnetic structure. The $z$-axis component
of $\mathbf{D}$ was never detected in \bacu, where it was assumed
to be weak. In Ref.~\onlinecite{Sato2002} it was tentatively
suggested that the out-of-plane component is dominant in the
Co-based $x=1$ material, and stabilizes a commensurate magnetic
structure. To verify this hypothesis and better understand the
underlying physics, a detailed knowledge of magnetic interactions
not only in \bacu\ ($x=0$), but also in \ba\ ($x=1$) is required.
In the present paper we report the results of neutron diffraction
and inelastic neutron scattering measurements on the $x=1$
material \ba.

\section{Experimental}
To date, the exact crystal structure of \ba\ has not been
determined. However, powder data\cite{lattice} indicate that the
material is very similar to its Cu-based counterpart and is
characterized by the $P\overline{4}2_1 m$ crystallographic space
group. The lattice parameters for \ba\ are $a=b=8.410$~\AA\ and
$c=5.537$~\AA, as measured at $T=10$~K. In each crystallographic
unit cell the magnetic Co$^{2+}$ ions are located at $(0,0,0)$ and
$(0.5,0.5,0)$ positions. The NN Co-Co distance is thus along the
$(1,1,0)$ direction and equal to $a/\sqrt{2}\approx 5.9$~\AA. For
the present study we utilized two single crystal samples prepared
using the floating zone technique. Both crystals were cylindrical,
roughly 5~mm diameter $\times$ 50~mm long, with a mosaic spread of
about $0.4^\circ$.

The first series of experiments was carried out at the HB1 3-axis
spectrometer installed at the High Flux Isotope reactor at Oak
Ridge National Laboratory (Setup I). Its main purpose was to
determine the spin arrangement in the magnetically ordered state.
The sample was mounted with the $b$ axis vertical making $(h,0,l)$
reflections accessible for measurements. Neutrons with a fixed
incident energy of 13.5~meV were used in combination with a
pyrolytic graphite (PG) monochromator and analyzer,
$30'-40'-20'-120'$ collimation, and a PG higher-order filter.
Sample environment was a closed-cycle refrigerator that allowed
measurements at temperatures down to 3.5~K.  To isolate the
magnetic contribution, integrated intensities were measured in a
series of rocking curves at $T=3.5$~K$<T_\mathrm{N}$ and
$T=10$~K$>T_\mathrm{N}$.
 \begin{figure}
 \includegraphics[width=3.2in]{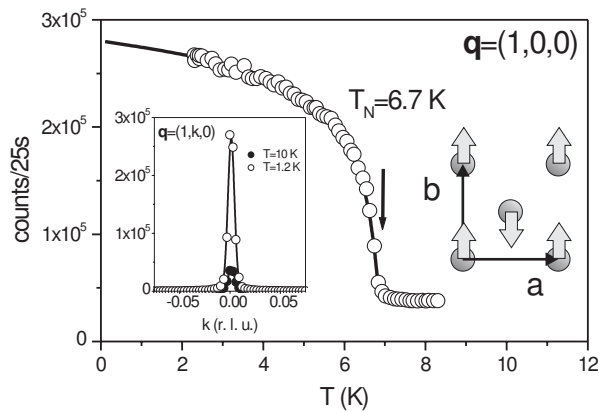}
 \caption{\label{orderp} Measured temperature dependence of the
$(1,0,0)$ magnetic Bragg peak intensity in \protect\ba\ (symbols).
The solid line is a guide for the eye. Left inset: transverse
scans (rocking curves) across the $(1,0,0)$ Bragg reflection
measured below (open circles) and above (solid circles) the N\'eel
temperature $T_N\approx6.7$~K. The residual intensity seen at
$T>T_\mathrm{N}$ is of non-magnetic origin and due to multiple
scattering. Right inset: the proposed model for the spin structure
of \ba. All spins are within the $(a,b)$ crystallographic plane.}
 \end{figure}
While using Setup I, it became apparent that the study of magnetic
excitations could be much better carried out using of a cold
neutron instrument, the relevant energy scale for \ba\ being about
2~meV. These measurements were therefore performed at the TASP
3-axis spectrometer installed at the SINQ spallation source at
Paul Scherrer Institut (Setup II). Neutrons with a fixed final
energy of 5.5~meV were used with PG monochromator and analyzer,
and a PG filter after the sample. The beam collimation was
$\mathrm{(guide)}-80'-80'-\mathrm{(open)}$. The sample was mounted
with the $c$ axis vertical, making momentum transfers in the
$(h,k,0)$ reciprocal-space plane accessible for measurement. Spin
wave dispersion curves were measured along the $(h,0,0)$ and
$(h,h,0)$ directions using constant-$Q$ scans in the energy range
0--4~meV. The sample environment was a standard ``ILL Orange''
He-4 flow cryostat, and most of the data were taken at $T=2$~K.

\section{Results}
\subsection{A model for the spin structure}
In the \bacu\ system magnetic ordering gives rise to
incommensurate peaks surrounding the integer $h$, $k$ and $l$
reciprocal-space points.\cite{Zheludev1996} In contrast, in \ba\
magnetic Bragg scattering was detected below $T_\mathrm{N}=6.7$~K
at strictly commensurate positions $h$, $k$ and $l$-integer. Due
to their location, the magnetic reflections, except those on the
$(h,0,0)$ and $(0,k,0)$ reciprocal-space rods, coincide with
nuclear ones. Figure~\ref{orderp} shows the measured temperature
dependence of the $(1,0,0)$ peak intensity (Setup II). The insert
shows rocking curves measured above and below the ordering
temperature. The appreciable residual intensity seen at $T>T_N$ at
the $(1,0,0)$ {\it forbidden} nuclear peak position is due to
multiple scattering. 19 non-equivalent magnetic Bragg intensities
measured using Setup I were normalized by the resolution volume,
which in a 3-axis experiment plays the role of the Lorentz factor.
In our case it was calculated using the Cooper-Nathans
approximation.
 The resolution-corrected magnetic intensities $I_\mathrm{obs}$ are
listed in Table~\ref{tab}.  The observed intensity pattern
indicates a planar spin arrangement, with all spins confined to
the $(a,b)$ plane, and nearest-neighbor spins aligned antiparallel
with each other.
 \begin{table}%[H] add [H] placement to break table across pages
 \caption{\label{tab} Magnetic Bragg intensities measured in \ba\
 at $T=3.5$~K in comparison to those calculated for
the proposed $(a,b)$-planar collinear antiferromagnetic
structure.}
 \begin{ruledtabular}
 \begin{tabular}{r r r r r r r}
$h$ & $k$& $l$ & $I_{\mathrm{calc}}$ & $I_{\mathrm{obs}}$ &
$\sigma_{\mathrm{obs}}$ &
$\frac{I_{\mathrm{obs}}-I_{\mathrm{calc}}}{\sigma_{\mathrm{obs}}}$\\
\hline
 -1 &   0 &   0 &     9031&     8226&      186&       -4.3\\
 -2 &   0 &   0 &        0&      100&      129&        0.8\\
 -3 &   0 &   0 &     5577&     5702&       84&        1.5\\
 -4 &   0 &   0 &        0&     1405&     4013&        0.4\\
 -5 &   0 &   0 &     2340&     3954&      212&        7.6\\
  0 &   0 &   1 &        0&      -22&     1323&       -0.0\\
 -1 &   0 &   1 &    13294&    12187&      230&       -4.8\\
 -2 &   0 &   1 &        0&      -59&      223&       -0.3\\
 -3 &   0 &   1 &     5880&      875&     1260&       -4.0\\
 -4 &   0 &   1 &        0&      220&      301&        0.7\\
 -5 &   0 &   1 &     2260&     3257&     1690&        0.6\\
  0 &   0 &   2 &        0&       21&      106&        0.2\\
 -2 &   0 &   2 &        0&       70&      243&        0.3\\
 -3 &   0 &   2 &     5018&     8775&      242&       15.5\\
 -4 &   0 &   2 &        0&     -150&      325&       -0.5\\
  0 &   0 &   3 &        0&      689&     3308&        0.2\\
 -1 &   0 &   3 &     5400&     5081&     1897&       -0.2\\
 -2 &   0 &   3 &        0&      269&      123&        2.2\\
 -3 &   0 &   3 &     3134&     4115&     6000&        0.2\\
 \end{tabular}
 \end{ruledtabular}
 \end{table}
 The alignment of nearest-neighbor spins along the
$c$ direction is ``ferromagnetic''. Such a spin structure is
identical to the one in the commensurate spin-flop phase of
\bacu\, stabilized by an external magnetic field applied along the
$c$ axis.\cite{ZheludevPRL1997,ZheludevPRB1998} As can be seen
from Table~\ref{tab}, where $I_\mathrm{calc}$ are the calculated
magnetic intensities, this simple collinear model reproduces our
limited diffraction data for \ba\ rather well. Due to the
possibility of antiferromagnetic domains, the spin orientation
{\it within} the $(a,b)$ plane could not be determined
unambiguously. Neither did we measure the actual magnitude of the
ordered moment, since the crystallographic data needed to bring
the measured magnetic intensities to an absolute scale is not
currently available for \ba. It is reasonable to assume that at
low temperatures the sublattice magnetization is close to its
classical saturation value. Indeed, this is the case in the
$S=1/2$ Cu$^{2+}$-system, where quantum fluctuations may be
expected to be even stronger than in the $S=3/2$ Co$^{2+}$
compound.

\subsection{Spin waves}

\begin{figure}
\includegraphics[width=3.2in]{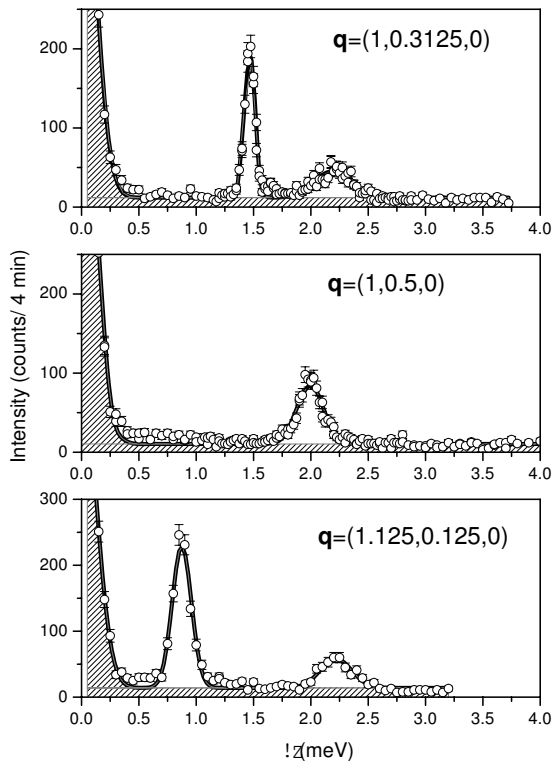}
\caption{\label{exdata} Typical constant-$Q$ scans measured in
\protect\ba\ at $T=2$~K. The solid lines are Gaussian fits to the
data. Shaded areas represent the background level.}
\end{figure}
The dispersion of spin wave excitations in \ba\ was found to be
quite different from that in \bacu.
\cite{ZheludevPRL1998,Zheludev1999} Figure~\ref{exdata} shows
typical constant-$q$ scans measured in the Co-compound using Setup
II at $T=2$~K. Two distinct sharp excitations are observed. One
branch is acoustic in origin, with excitation energy linearly
going to zero at the AF zone-center $(1,1,0)$. The second
``optical'' branch is barely dispersive and is always seen around
2~meV energy transfer. The two spin wave branches converge at the
AF zone-boundary $(0.5,0.5,0)$. In all cases the observed energy
width of spin wave peaks is resolution limited. The apparent
variation of peak width seen in Fig.~\ref{exdata} is due to
instrumental ``focusing'' effects. The data were analyzed using
Gaussian fits (solid line in Fig.~\ref{exdata}). The background
(shaded areas) was assumed to be constant with an additional
Gaussian component at zero energy transfer to model incoherent
elastic scattering. The dispersion relations along the $(1,1,0)$
and $(1,0,0)$ reciprocal-space directions deduced from these fits
are plotted in symbols Fig.~\ref{disp}.

\subsection{Data analysis}
\begin{figure}
\includegraphics[width=3.2in]{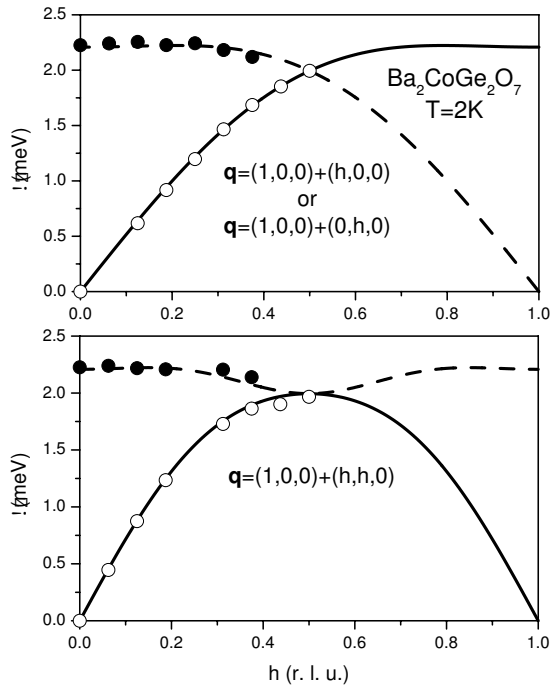}
\caption{\label{disp} Dispersion of spin waves along the $(1,0,0)$
and $(1,1,0)$ reciprocal-space directions measured in \protect\ba\
at $T=2$~K (symbols). The solid and dashed lines represent the two
spin wave branches in the model defined by Eq.~\protect\ref{ham},
with dispersion relations given by Eq.~\protect\ref{disprel} and
parameters chosen to best-fit the data.}
\end{figure}
The observed dispersion of spin wave excitations can be understood
in the framework of linear spin wave theory. To construct a model
spin Hamiltonian, we assumed that the dominant magnetic
interactions in \ba\ are those between nearest-neighbor spins in
the $(a,b)$ plane, as is the case in \bacu.  Given that the
$S=3/2$ Co$^{2+}$ ions are frequently associated with a large
magnetic anisotropy, in our model we allowed this coupling to be
anisotropic, and also included a single-ion anisotropy term. The
resulting model Hamiltonian is written as:
\begin{eqnarray}
 \hat{\mathcal{H}} & = &
 \sum_m\tilde{\sum_n} \left[J_z S_m^{(z)}S_n^{(z)}+J_{\bot}S_m^{(x)}S_n^{(x)}+ J_{\bot}S_m^{(y)}S_n^{(y)}\right] + \nonumber \\
  & + & A\sum_m \left[S_m^{(z)}\right]^2+A\sum_n
  \left[S_n^{(z)}\right]^2.\label{ham}
 \end{eqnarray}
Here $m$ and $n$ label the spins on the two antiferromagnetic
square sublattices with origins at $(0,0,0)$ and $(0.5,0.5,0)$,
respectively, and $\tilde{\sum}$ stands for summation over nearest
neighbors. Note that the interactions along the $c$ axis are not
included in the above expression. They were not measured directly
in this work, and are likely to be ferromagnetic, due to the value
of the magnetic ordering vector. The corresponding coupling
constant was previously found to be extremely weak in
\bacu,\cite{Zheludev1996} and was assumed to also be small in the
Co-system. The $c$-axis coupling should thus have no considerable
effect on spin wave dispersion in the $(h,k,0)$ reciprocal-space
plane.

The spin wave Hamiltonian is obtained from Eq.~\ref{ham} through a
Holstein-Primakoff transformation. After linearization it can be
easily diagonalized by a Fourier-Bogolyubov transformation. This
straightforward yet tedious calculation for our particular case
yields two spin wave branches with the following dispersion
relations:
\begin{equation}
 [\hbar \omega_{\mathbf{q}}^{(\pm)}]^2  = S^2 ( 8J_\bot\mp4J_\bot
 C_{\mathbf{q}})( 8J_\bot\pm4J_z
 C_{\mathbf{q}}+2A),\label{disprel}
\end{equation}
where
\begin{equation}
 C_{\mathbf{q}} \equiv \cos(\pi h+ \pi k)+\cos(\pi h- \pi k).
\end{equation}
From Eq.~\ref{disprel} it follows that exchange anisotropy and
single-ion anisotropy can not be distinguished based on dispersion
measurements alone. Indeed, it can be rewritten using only two
independent parameters:
\begin{equation}
 [\hbar \omega_{\mathbf{q}}^{(\pm)}]^2  = (4JS)^2 (2\mp
 C_{\mathbf{q}})( 2\mathcal{A}\pm
 C_{\mathbf{q}}),\label{disprel2}
\end{equation}
where $J\equiv\sqrt{J_{\bot}J_z}$ and $\mathcal{A}\equiv
\frac{J_\bot}{J_z} +\frac{A}{4J_z}$. The quantity $\mathcal{A}$ is
unity in the isotropic case and is a generalized measure of
easy-plane anisotropy in our model. Excellent fits to the data are
obtained assuming $S=3/2$ and using $J=0.103(1)$~meV and
$\mathcal{A}=2.58(3)$. Dispersion curves calculated using these
parameters are shown in dashed and solid lines in Fig.~\ref{disp}.
Unlike its Cu-based counterpart, \ba\ is characterized by {\it
very strong magnetic easy-plane anisotropy}.

\section{Discussion}
The strong anisotropy effects in \ba\ push all spins in the system
into the $(a,b)$ crystallographic plane. This effect is similar to
that of a magnetic field applied along the $c$ axis that favors an
$(a,b)$-planar state in \bacu.\cite{ZheludevPRL1998,Zheludev1999}
Since the helimagnet-forming uniform component $\mathbf{D}_{xy}$
of the Dzyaloshinskii vector is itself in the $(a,b)$ plane,
forcing the spins into the $(a,b)$ plane makes the corresponding
triple-product in the Hamiltonian $\mathbf{D}_{xy}(\mathbf{S}_m
\times \mathbf{S}_n)$ vanish. Only the non-helimagnet-forming
sign-alternating $z$-axis component of $\mathbf{D}$ remains
relevant. As a result, the spin structure may be slightly canted,
but is, nevertheless, commensurate.

It is important to stress that effective easy-plane anisotropy was
previously detected in \bacu\ as well. However, in this $S=1/2$
Cu-based system any single-ion term is reduced to a constant and
is therefore irrelevant. The only source of anisotropy is a
two-ion term, which was shown to be caused by the so-called KSEA
interactions.\cite{ZheludevPRL1998,Zheludev1999} The latter are a
very weak effect with an energy scale of $D^2/J\sim 3\cdot 10^{-2}
J$. On a square lattice KSEA interactions happen to be just strong
enough to distort the helical structure, but not to fully destroy
incommensurability.\cite{ZheludevPRL1998} In contrast, as follows
from the present study, easy-plane anisotropy in \ba\ is {\it much
stronger}, of the order of $J$ itself. The anisotropy is probably
due to single-ion effects that are only allowed for $S>1/2$, and
its magnitude is well beyond the critical value needed to destroy
the helimagnetic state.

The results discussed above allows us to speculate about the IC
transition in \bacuco\ that occurs with increasing
Co-concentration $x$. Each Co-impurity {\it strongly} ``pins'' the
original spiral at the impurity site, firmly confining the
corresponding spin to the $(a,b)$ plane. Helimagnetic correlations
are totally destroyed when the characteristic distance between
such strong-pinning locations becomes comparable with the period
of the unperturbed spiral, which in \bacu\ is roughly 40
nearest-neighbor bonds.\cite{Zheludev1996} This suggests a
critical concentration of about $x\approx 2$\%, in reasonable
agreement with bulk magnetization data of
Ref.~\onlinecite{Sato2002}.

\section{Conclusion}
To summarize, the commensurate nature of the ground state in \ba\
is primarily due not to a dominant staggered component of the
Dzyaloshinskii vector, but to easy-plane anisotropy effects that
are {\it orders of magnitude stronger} than typical
Dzyaloshinskii-Moriya or KSEA interactions. As a result, the
destruction of helimagntism in \bacuco\ occurs very rapidly with
increasing Co-concentration, as soon as the mean distance between
impurities becomes comparable to the period of the spin spiral.

\begin{acknowledgments}
This work is supported in part by the Grant-in-Aid for COE
Research ``SCP coupled system" of the Ministry of Education,
Culture, Sports, Science and Technology, Japan. Work at ORNL and
BNL was carried out under DOE Contracts No. DE-AC05-00OR22725 and
DE-AC02-98CH10886, respectively.
\end{acknowledgments}

%\bibliographystyle{apsrev}
%\bibliography{mag}

\end{document}